\documentclass[aps,prl,twocolumn,superscriptaddress,showpacs,preprintnumbers,amsmath,amssymb]{revtex4}

\usepackage{graphicx} 
\usepackage{dcolumn}  

\graphicspath{{ps}}

\def\simle{\mathrel{\rlap{\raise 0.511ex \hbox{$<$}}{\lower 0.511ex \hbox{$\sim$}}}}

\begin{document}

\title{\large
Search for
$D^0-\overline{D}{}^0$ mixing using semileptonic decays at Belle}

\affiliation{Budker Institute of Nuclear Physics, Novosibirsk}
\affiliation{Chiba University, Chiba}
\affiliation{Chonnam National University, Kwangju}
\affiliation{University of Cincinnati, Cincinnati, Ohio 45221}
\affiliation{University of Frankfurt, Frankfurt}
\affiliation{University of Hawaii, Honolulu, Hawaii 96822}
\affiliation{High Energy Accelerator Research Organization (KEK), Tsukuba}
\affiliation{Hiroshima Institute of Technology, Hiroshima}
\affiliation{Institute of High Energy Physics, Chinese Academy of Sciences, Beijing}
\affiliation{Institute of High Energy Physics, Vienna}
\affiliation{Institute for Theoretical and Experimental Physics, Moscow}
\affiliation{J. Stefan Institute, Ljubljana}
\affiliation{Kanagawa University, Yokohama}
\affiliation{Korea University, Seoul}
\affiliation{Swiss Federal Institute of Technology of Lausanne, EPFL, Lausanne}
\affiliation{University of Ljubljana, Ljubljana}
\affiliation{University of Maribor, Maribor}
\affiliation{University of Melbourne, Victoria}
\affiliation{Nagoya University, Nagoya}
\affiliation{Nara Women's University, Nara}
\affiliation{National Central University, Chung-li}
\affiliation{National United University, Miao Li}
\affiliation{Department of Physics, National Taiwan University, Taipei}
\affiliation{H. Niewodniczanski Institute of Nuclear Physics, Krakow}
\affiliation{Nippon Dental University, Niigata}
\affiliation{Niigata University, Niigata}
\affiliation{Nova Gorica Polytechnic, Nova Gorica}
\affiliation{Osaka City University, Osaka}
\affiliation{Osaka University, Osaka}
\affiliation{Panjab University, Chandigarh}
\affiliation{Peking University, Beijing}
\affiliation{Princeton University, Princeton, New Jersey 08544}
\affiliation{University of Science and Technology of China, Hefei}
\affiliation{Seoul National University, Seoul}
\affiliation{Shinshu University, Nagano}
\affiliation{Sungkyunkwan University, Suwon}
\affiliation{University of Sydney, Sydney NSW}
\affiliation{Tata Institute of Fundamental Research, Bombay}
\affiliation{Toho University, Funabashi}
\affiliation{Tohoku Gakuin University, Tagajo}
\affiliation{Tohoku University, Sendai}
\affiliation{Department of Physics, University of Tokyo, Tokyo}
\affiliation{Tokyo Institute of Technology, Tokyo}
\affiliation{Tokyo Metropolitan University, Tokyo}
\affiliation{Tokyo University of Agriculture and Technology, Tokyo}
\affiliation{University of Tsukuba, Tsukuba}
\affiliation{Virginia Polytechnic Institute and State University, Blacksburg, Virginia 24061}
\affiliation{Yonsei University, Seoul}
   \author{U.~Bitenc}\affiliation{J. Stefan Institute, Ljubljana} 
   \author{K.~Abe}\affiliation{High Energy Accelerator Research Organization (KEK), Tsukuba} 
   \author{K.~Abe}\affiliation{Tohoku Gakuin University, Tagajo} 
   \author{I.~Adachi}\affiliation{High Energy Accelerator Research Organization (KEK), Tsukuba} 
   \author{H.~Aihara}\affiliation{Department of Physics, University of Tokyo, Tokyo} 
   \author{Y.~Asano}\affiliation{University of Tsukuba, Tsukuba} 
   \author{T.~Aushev}\affiliation{Institute for Theoretical and Experimental Physics, Moscow} 
   \author{S.~Bahinipati}\affiliation{University of Cincinnati, Cincinnati, Ohio 45221} 
   \author{S.~Banerjee}\affiliation{Tata Institute of Fundamental Research, Bombay} 
   \author{E.~Barberio}\affiliation{University of Melbourne, Victoria} 
% ...
\collaboration{The Belle Collaboration}

\date{\today}

\begin{abstract}

A search for mixing in the neutral $D$ meson system has been performed using 
semileptonic $D^0\to K^{(*)-}e^+\nu$ decays. Neutral $D$ mesons from
$D^{\ast+}\to D^0\pi^+$ decays are used;  the flavor at production is
tagged by the charge of the slow pion. 
The measurement is performed using 
253\,fb$^{-1}$ of data recorded by the Belle detector. 
From the yield of right-sign and wrong-sign decays arising from
non-mixed and mixed events, respectively, we estimate the upper limit of
the time-integrated mixing rate to be $r_D < 1.0\times 10^{-3}$ at $90\%$ C.L.

\end{abstract}


\maketitle
\tighten

{\renewcommand{\thefootnote}{\fnsymbol{footnote}}}
\setcounter{footnote}{0}


\centerline{\bf I. INTRODUCTION}
~

The phenomenon of mixing has been observed in the $K^0-\overline{K}{}^0$
and $B^0-\overline{B}{}^0$ systems, but not yet in the $D^0-\overline{D}{}^0$
system. 
The parameters used to characterize $D^0-\overline{D}{}^0$ mixing are $x = \Delta m /
\overline{\Gamma}$ and $y = \Delta \Gamma / 2\overline{\Gamma}$, where
$\Delta m$ and $\Delta \Gamma$ are the differences in mass and decay
width between the two neutral charmed meson mass eigenstates, and
$\overline \Gamma$ is the mean decay width. The mixing rate 
within the Standard Model is expected to be small \cite{small_mix}:
the largest predicted values, including the impact of long
distance dynamics, are of order 
$x \simle y \sim
10^{-3}-10^{-2}$. Observation of a mixing rate significantly larger than
predicted would indicate either new physics (enhanced $x$) or
insufficient understanding of long distance effects (larger $y$).

For $x, y \ll 1$ and negligible CP violation, the time-dependent mixing
probability for semileptonic $D^0$ decays is \cite{Xing}
\begin{equation}
{\cal{P}}(D^0\to\overline{D}{}^0\to X^+\ell^-\overline{\nu}_\ell)\propto r_D
~t^2~e^{-\Gamma t},
\label{eq_time_dep}
\end{equation}
where $r_D$ is the ratio of the time-integrated mixing 
to the time-integrated non-mixing probability:

\begin{equation}
r_D
= {\int_0^\infty
  dt~{\cal{P}}(D^0\to\overline{D}{}^0\to X^+\ell^-\overline{\nu}_\ell)\over 
\int_0^\infty dt~{\cal{P}}(D^0\to X^-\ell^+\nu_\ell)}
\approx{x^2+y^2\over 2}.
\label{eq3}
\end{equation} 

In this paper we present a search for $D^0-\overline{D}{}^0$ mixing 
using semileptonic decays of charmed mesons. 
The flavor of a neutral $D$ meson at production is determined from 
the charge of the accompanying slow pion ($\pi_s$) in the $D^{*+} \to D^0
\pi_{\rm s}^+$ decay \cite{charge}. We reconstruct the $D^0$ as $D^0 \to
K^{(*)-} e^+ \nu_e$   
and identify its flavor at the time of decay from the charge correlation of
the kaon and the electron. We search for mixing by
reconstructing the ``wrong sign'' (WS) decay chain,  
$D^{*+} \to D^0 \pi_{\rm s}^+$, $D^0 \to \overline{D}{}^0$, 
$\overline{D}{}^0 \to K^{(*)+} e^- \overline{\nu}_e$, which results in a WS
charge combination of the three particles used to reconstruct the
candidate. 
The non-mixed process results in a ``right sign'' (RS) charge
combination, $\pi_{\rm s}^+ K^- e^+$.  
In contrast to hadronic decays, the WS charge combinations
can occur only through mixing,
and $r_D$ can be obtained directly as the ratio of WS to RS signal events.

We make no attempt to reconstruct $K^{*-}$ mesons. We treat
charged daughter particles  
 ($K^-$ or $\pi^-$)
as if they were the direct daughters of the $D^0$, omitting the
accompanying neutral particle. 
A small contribution from $D^0 \to \rho^- e^+ \nu_e$ and  $D^0 \to
\pi^- e^+ \nu_e$ decays, due to charged pions surviving the kaon
selection, is included in the analysis.

We measure $r_D$ in a 253\,$\rm{fb}^{-1}$ data sample
recorded by the Belle detector at
the KEKB asymmetric-energy $e^+e^-$ collider \cite{pospesevalnik}, at a
center-of-mass (cms) energy of about 10.6\,GeV.  
The Belle detector \cite{detektor} is a large-solid-angle magnetic spectrometer that
consists of a silicon vertex detector (SVD), a 50-layer central drift
chamber (CDC), an array of aerogel threshold \v{C}erenkov counters
(ACC), a barrel-like arrangement of time-of-flight scintillation
counters (TOF), and an electromagnetic calorimeter (ECL) comprised of
CsI(Tl) crystals located inside a superconducting solenoid coil that
provides a $1.5$\,T magnetic field. An iron flux-return located outside
of the coil is instrumented to detect $K^0_L$ mesons and to identify
muons (KLM). 
 Two different inner detector configurations were
used. 
The first 140\,$\rm{fb}^{-1}$ of data were taken using a 2.0\,cm
radius beampipe and a 3-layer silicon vertex detector, and the subsequent 
113\,$\rm{fb}^{-1}$ were taken using a $1.5$\,cm radius beampipe, a 
4-layer silicon detector and a small-cell inner drift chamber \cite{Belle2}.

Simulated events are generated by the QQ generator and processed with
a full simulation of the Belle detector, using the GEANT package
\cite{qq98}. 
To simulate mixed $D$
meson decays we use generic (non-mixed) Monte Carlo (MC) events and 
 appropriately reweight the proper decay time distribution.

~\\
~\\
\centerline{\bf II. SELECTION AND RECONSTRUCTION}
~

For $\pi_{\rm s}^+$ candidates we consider tracks with: momentum $p < {\rm
600\,MeV}/c$; projections of the impact parameter with respect to the
interaction point in the radial and beam
directions $dr<1$\,cm and $|dz|<2$\,cm, respectively;
and electron identification likelihood ratio, based on
the information from the CDC, ACC and ECL \cite{e_id}, ${\cal L}_e <
0.1$. 
Electron candidates are required to have $p>600$\,MeV/$c$ and ${\cal
  L}_e > 0.95$. 
Kaon candidates are chosen from the remaining tracks in the event with
$p>$\,800\,MeV/$c$ and
${\cal{L}}(K^\pm)/({\cal{L}}(K^\pm)+{\cal{L}}(\pi^\pm))>0.5$,
where $\cal{L}(K^\pm,\pi^\pm)$ is the likelihood that a given track is a 
$K^\pm$ or a $\pi^\pm$ based on information from the TOF, CDC
and ACC. Charged pions originating from
semileptonic $D^0\to\pi^-e^+\nu_e$ decays that pass the kaon selection
criteria are treated as kaons and assigned the kaon mass.
According to MC simulation, this  selection retains about 58\% (43\%,
52\%) of signal slow pions (electrons, kaons), and at this stage about
13\% (5\%, 44\%) of the selected $\pi_s$ ($e$, $K$) candidates are
misidentified.

The variable used to isolate the signal events is 
$\Delta M = M(\pi_{\rm s} K e \nu_e)-M(K e \nu_e)$,
where $M(\pi_{\rm s} K e \nu_e)$ and $M(K e \nu_e)$ are the invariant
masses of the selected charged particles and reconstructed neutrino
(see below), with and without the slow pion.

We require the cms momentum of the kaon-electron system to be $p_{\rm
cms}(Ke)>2$\,GeV/$c$ to reduce the combinatorial background and the
background from $B\overline{B}$ events, and to improve the 
resolution in $\Delta M$. To further suppress the contribution of
 $B\overline{B}$ events, we require the ratio of
the second to the zeroth Fox-Wolfram moments, known as $R_2$, to be
greater than 0.2 \cite{fox-wolfram}.

Background from $D^0 \to K^- \pi^+$ decay
is suppressed by requiring the invariant mass of the $K$-$e$ system to
satisfy $M(Ke) < 1.82$\,GeV/$c^2$.
Background from $D^0\to K^-K^+$ decay
is effectively
reduced by the requirement $|M(KK)-m_{D^0}|>10$\,MeV/$c^2$,
where $M(KK)$ is the  $K$-$e$ invariant mass, calculated using the kaon
mass for both tracks.

The background from photon conversions, where the $e^-$ and $e^+$ are
taken to be the electron and slow pion candidates, is
suppressed by requiring $M(e^+e^-) >
150$\,MeV/$c^2$, where $M(e^+e^-)$ is the invariant mass of the $\pi_{\rm
s}^+$-$e^-$ system with the electron mass assigned to the pion.
A further search for an additional $e^\pm$ with charge opposite
to that of the electron (slow pion) candidate in the $D^0$ ($D^{*+}$)
decay is performed in each event. Again, candidates with $M(e^+e^-)$ below
150\,MeV/$c^2$ are rejected.

The first approximation to the neutrino four-momen\-tum is calculated from
four-momentum conservation: 
\begin{equation}
P_\nu=P_{\rm cms}-P_{\pi_{\rm s} K e} -P_{\rm rest},
\label{neutrino}
\end{equation}
where $P_{\rm cms}$ denotes the cms four-momentum of the $e^+e^-$
system, $P_{\pi_{\rm s} K e}$ the cms four-momentum of the $\pi_{\rm
  s}$-$K$-$e$ combination, and $P_{\rm rest}$ the cms four-momenta of all
remaining charged tracks and photons in the event.
The $\Delta M$ resolution is significantly improved by applying two
kinematic constraints to correct $P_{\rm rest}$. First, 
events with
$-4\,\mathrm{GeV}^2/c^4~<M(\pi_{\rm s} K e\nu)^2<36\,\mathrm{GeV}^2/c^4$
are selected and 
the magnitude of $P_{\rm rest}$ is rescaled by a factor $x$ such that
$M(\pi_{\rm s} K e\nu)^2=(P_{\rm cms} - x P_{\rm rest})^2\equiv
m_{D^{\ast \pm}}^2$, 
with $m_{D^{\ast \pm}}$ fixed to its nominal value \cite{PDG}. 
Next, 
only events with $-2\,\mathrm{GeV}^2/c^4\,<M_\nu^2<0.5\,\mathrm{
GeV}^2/c^4$ are retained and
the direction of the three-momentum ${\vec{p}_{\rm rest}}$ is corrected 
in order to yield $M_\nu^2 \equiv 0$.
The correction rotates $\vec{p}_{\rm rest}$
in the plane determined by the vectors $\vec{p}_{\rm rest}$ and $\vec{p}_{\pi_{\rm s} K e}$.

The mass difference $\Delta M$ is then calculated using this reconstructed
neutrino four-momentum.  
The full width at half maximum (FWHM) of the $\Delta M$ peak in MC simulated 
$D^0 \to K^- e^+ \nu$ events is found to be 
7\,MeV/$c^2$. Events with 
$\Delta M < 0.18$\,GeV/$c^2$ are retained for further analysis.

According to MC simulation, the $D^0 \to K^{*-} e^+ \nu_e$,
$\rho^- e^+ \nu_e$, and $\pi^- e^+ \nu_e$ modes have 
$\Delta M$ distributions similar to that from $D^0 \to K^- e^+ \nu_e$ decay,
but with a larger FWHM. 
The charge correlation is preserved if the slow pion decays into
a muon, and the muon is taken as the slow pion candidate. 
These decays 
will be referred to as ``associated signal'' decays. Their fraction in
the reconstructed MC signal sample is $14.9 \pm 0.1\%$,
with the quoted error being due to MC statistics only.

According to MC, the backgrounds for both RS and WS decays consist
mainly of combinatorial background (98.4\% and 98.0\%, respectively). 
In order to minimize systematic uncertainties, we model the shape of
this background using data, following a method validated with MC events.
We use candidates with the unphysical sign combinations $\pi_{\rm s}^+ K^-
e^-$ and $\pi_{\rm s}^+ K^+ e^+$ to model the combinatorial background
in the RS sample. In the WS sample this background 
is mainly due to random $\pi_{\rm s}^+$ candidates combined with the
selected kaon and electron candidates, and
is modelled by combining $\pi_{\rm s}^+$ and ($K^-+e^+$) candidates
from different events.
The correlated background (1.6\% in the RS, 2.0\% in the WS sample)
arises from a true $\pi_{\rm s}^+$ from a $D^{*+}$ decay combined with
true or misidentified $K$ and $e$ candidates, both (in the RS sample) or at
least one of them (WS sample) from the corresponding $D^0$ decay.
This background is described using MC.

\begin{figure}[h]
\includegraphics[width=0.4\textwidth]{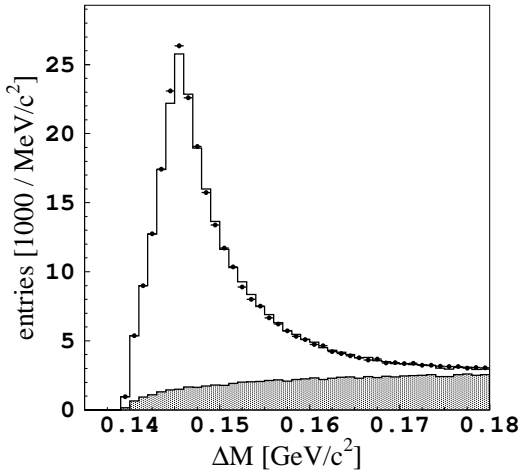}
\caption{ $\Delta M$ distribution for RS events.  The data
  (background) is represented by points (hatched
  histogram), and the result of the binned maximum likelihood fit by the
  solid line. }
\label{fig_dm_RS}
\end{figure}

Figure\,\ref{fig_dm_RS} shows the $\Delta M$ distribution for RS events.  
We perform a binned 
maximum likelihood fit to the distribution, maximizing 
\begin{equation}
{\cal L} = \prod_{i=1}^{N_{\rm bin}}\frac{e^{-{\mu}(\Delta M_i)}\times
  {(\mu}(\Delta M_i))^{N_i}}{N_i !},
\label{eq_fit}
\end{equation}
where $N_i$ is the number of entries in the $i$-th bin and $\mu(\Delta
M_i)$ is the expected number  
of events in this bin, given by
\begin{equation}
{\mu}(\Delta M_i)={\cal{N}}[f_s P_s(\Delta M_i)+(1-f_s) P_b (\Delta M_i)].
\label{eq_predicted}
\end{equation}
$P_s$ ($P_b$) is the signal (background) $\Delta M$ distribution
obtained from MC (as described above); $f_s$ and $\cal{N}$ are
the signal fraction and the overall normalization, respectively, and are free
parameters in the fit. The hatched histogram in Fig.\,\ref{fig_dm_RS}
shows the fitted background contribution. The fitted signal fraction
is $f_s = 73.4 \pm 0.1\%$ and the number of RS signal events is
$N_{\rm RS}^{\rm tot} = 229\,452\pm 597$.

We increase the sensitivity to $D^0-\overline{D}{}^0$ mixing by exploiting the
measurement of the $D^0$ proper decay time. 
The decay time is evaluated using the measured momentum of
the meson and the distance from the $e^+e^-$ interaction point to the
reconstructed $D^0$ decay $K$-$e$ vertex.  Due to the shape of
the KEKB accelerator interaction region, which is narrowest in the
vertical ($y$) direction, the dimensionless proper decay time $t_y$ is
calculated as
\begin{equation}
t_y={m_{D^0}\over c\tau_{D^0}}~{y_{\rm vtx}-y_{\rm
IP}\over p_y},
\label{eq8}
\end{equation}
where $p_y$ is the $y$ component of the $D^0$ candidate's momentum and $y_{\rm
vtx}$ and $y_{\rm IP}$ are the $y$ coordinates of the reconstructed
$K$-$e$ vertex and of the interaction point, respectively. 
$m_{D^0}$ and $\tau_{D^0}$ are the nominal mass and lifetime of $D^0$
mesons \cite{PDG}.

Mixed and non-mixed processes have different proper decay time
distributions: $t^2 e^{-t/\tau_{D^0}}$ and
$e^{-t/\tau_{D^0}}$, respectively.  
In order to calculate the mixing parameter $r_D$ after 
any selection based on
$D^0$ proper decay time, the ratio of the two types of
signal events, $N_{\rm WS}/N_{\rm RS}$, must be multiplied by the ratio of
their efficiencies, $\epsilon_{\rm RS}/\epsilon_{\rm WS}$.
The efficiencies are obtained by integrating the convolution of the above proper
decay time distributions with the detector resolution function, over
the selected $t_y$ interval.
As the observed proper decay time distribution is a convolution
of the signal and background probability density functions (p.d.f.) with the
detector resolution function, the latter is found by performing
a binned $\chi^2$ fit to the RS event $t_y$ distribution:
$\int_0^\infty dt \bigl[f_s  
e^{-t/\tau_{D^0}} + (1-f_s)(f_e e^{-t/\tau_{bkg}} +
(1-f_e)\delta(t))\bigr]\times{\cal R}(t/\tau_{D^0}-t_y)$. The signal
fraction $f_s$  
is obtained from the fit to the $\Delta M$ distribution. 
The fraction $f_e$ of the non-prompt background component and its
lifetime $\tau_{bkg}$  
are fixed to the values obtained by
fitting the MC background $t_y$ distributions; $\delta (t)$
describes the shape of the prompt background component. 
The resolution function ${\cal R}(t-t_y)$ is described phenomenologically 
by the sum of three Gaussians and an additional term for badly 
reconstructed tracks (``outliers''); we obtain the widths and
coefficients from the fit. 

The ratios $\epsilon_{\rm RS}/\epsilon_{\rm WS}$  are given in Table \ref{tab_t_bins}.
The errors are obtained by varying each parameter in the proper decay time 
fit by $\pm 1\sigma$, repeating the fit and recalculating the ratios;
the resulting changes are summed in quadrature. 
Using this method, the majority of systematic errors due to the
imperfect description of the decay time distribution cancel out. 

Comparison of the expected $t_y$ distribution for WS signal and
background events indicates that the figure-of-merit,
defined as $N_{\rm WS}^{\rm sig}/\sqrt{N_{\rm WS}^{\rm bkg}}$, is
optimal for $t_y\ge 1.0$.  Events that satisfy this condition
are retained for further analysis.

~\\
~\\
\centerline{\bf III. FIT AND RESULTS}
~

\begin{figure}[h]
\includegraphics[width=0.5\textwidth]{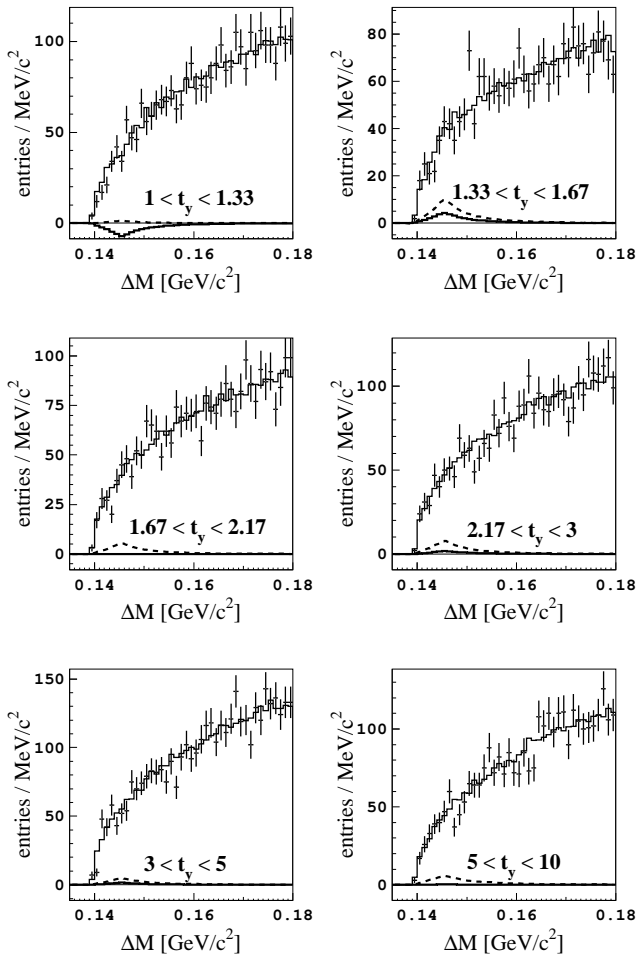}
\caption{ WS $\Delta M$ distributions in six proper decay time
intervals for data (points with error bars) and the results of the fit
described in the text (solid line). At the bottom of each figure the
fitted (expected for $r^i_D$ = 90\% C.L. upper limit) signal yield is
plotted as a solid (dashed) line. } 
\label{fig_dm_WS}
\end{figure}

\begin{figure}[h]
\includegraphics[width=0.5\textwidth]{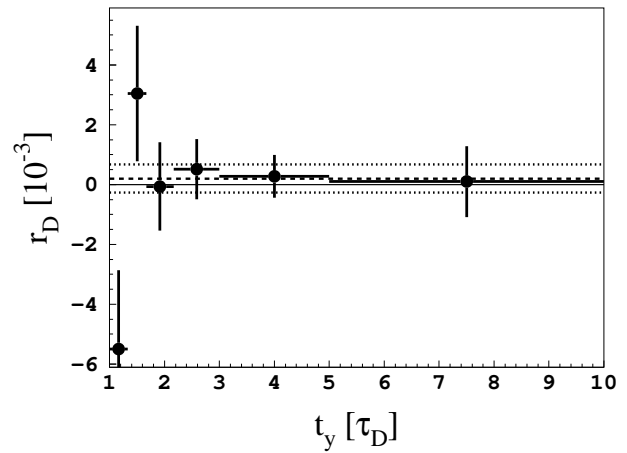}
\caption{ Measured $r_D^i$ values (points with error bars) in six
  different proper decay time intervals. The solid (dashed) line 
shows the null value (the fit to the six $r_D^i$ using a constant).
  The dotted lines denote the statistical error of the fit.} 
\label{fig_rD_tbins}
\end{figure}

We extract the RS and WS signal yields separately in six
intervals of $t_y$. The yields are obtained from binned maximum
likelihood fits to $\Delta M$, described by Eq.~\ref{eq_fit}; the WS
signal distribution is taken to be the same as the RS.
The results are shown in Table I and Fig.~\ref{fig_dm_WS}.
We obtain the time-integrated mixing probability ratio in the $i$-th $t_y$
interval, 
$r^i_D =N^i_{\rm WS}/N^i_{\rm RS} \times \epsilon_{\rm 
RS}^i/\epsilon_{\rm WS}^i$, by multiplying the ratio of WS to RS
signal events in each interval by the $t_y$ efficiency ratio. 
The mixing probability is compatible with zero in all proper decay
time intervals (see Fig.~\ref{fig_rD_tbins}).

\begin{table}[h]

\caption{The number of fitted signal events in the RS and WS samples,
  the efficiency ratio
$\epsilon_{\rm RS}^i/\epsilon_{\rm WS}^i$, and
the resulting $r_D^i$ value for each proper decay time interval.}
\begin{tabular}{c c c c c}

\hline
\hline
$t_y$ 			&$N^i_{\rm RS}$	&	$N^i_{\rm WS}$	&
$\displaystyle{{\epsilon_{\rm RS}^i}/ \epsilon_{\rm WS}^i}$	&	$r_D^i~[10^{-3}]$		\\
\hline
1.00-1.33	& 18\,742$\pm$166 &  $-$63.7$\pm$30.2	& 1.62$\pm$0.11 & $-$5.49$\pm$2.63 \\
1.33-1.67	& 15\,032$\pm$147 &  \phantom{$-$}40.3$\pm$29.9	& 1.14$\pm$0.08	& \phantom{$-$}3.05$\pm$2.27	\\
1.67-2.17	& 16\,430$\pm$155 &  \phantom{$4$}$-$1.3$\pm$30.6	& 0.79$\pm$0.05	&$-$0.06$\pm$1.48	\\
2.17-3.00	& 16\,691$\pm$157 &  \phantom{$-$}17.0$\pm$33.1	& 0.51$\pm$0.03	& \phantom{$-$}0.51$\pm$1.00	\\
3.00-5.00	& 15\,443$\pm$160 &  \phantom{$-$}14.7$\pm$37.4	& 0.30$\pm$0.01	& \phantom{$-$}0.28$\pm$0.72	\\
5.00-10.0	& \phantom{$1$}8\,263$\pm$123 &   \phantom{$-1$}3.0$\pm$35.2 &
0.28$\pm$0.02	& \phantom{$-$}0.10$\pm$1.19	\\
\hline
\hline

\end{tabular}

\label{tab_t_bins}
\end{table}

The overall $r_D$ follows from a $\chi^2$ fit with a constant to the
measured $r_D^i$ values, using a constant: 

\begin{equation}
r_D = (0.20 \pm 0.47)\times 10^{-3}.
\label{eq_res1}
\end{equation}

\noindent The quoted error is statistical only.

~\\
~\\
\centerline{\bf IV. SYSTEMATIC UNCERTAINTIES}
~

The experimental procedure was checked using a dedicated MC
sample of mixed $D^0$ decays. These were added in different
proportions to the generic MC, which includes both non-mixed $D^0$
decays and all known types of background decays. 
The application of
the same method used for the data on these samples verified 
that the
reconstructed $r_D$ value reproduces the input value without any
significant bias.

The main source of systematic error is the limited statistics of
the fitting distributions, predominantly the background distribution in
the WS sample. To estimate this uncertainty, we vary the contents of
all bins of the RS and WS $P_s(\Delta M)$ and $P_b(\Delta M)$
distributions independently in accordance with each bin's statistical uncertainty,
repeat the fit to the RS and WS data, calculate the corresponding $r_D^i$ 
in each proper decay time interval, and obtain a new $r_D$
value.
Repeating the procedure, the obtained distribution of
$r_D$ values has a Gaussian shape with a width of
$0.12 \times 10^{-3}$, which is taken as the systematic error due
to the limited statistics of the fitting distributions.

To estimate the systematic uncertainty due to the $\Delta M$  binning
and the shape
of the WS background, the $\Delta M$ distributions in each 
proper decay time interval, originally divided into 
45 bins in the range $0.135\,{\rm GeV}/c^2 < \Delta M < 0.180$\,GeV/$c^2$, 
are re-binned into 12, 15, 20, 30 and 60 bins in the same range. 
The fitting procedure is repeated for each set of bins, and the
systematic error due to binning is taken as half the
difference between the largest and the smallest $r_D$: $\pm 0.07 \times
10^{-3}$. 

According to MC, the majority of the correlated WS background arises
from $D^0\to K^-\pi^+\pi^0$ decays. We repeat the WS 
fits, changing the amount of this background in accordance with the
relative error on $\mathcal{B}(D^0\to K^-\pi^+\pi^0)$. The resulting variation
in $r_D$ is $\pm 0.02\times 10^{-3}$.
The same method applied in the fit to the RS sample introduces a
negligible variation of the $r_D$ value (less than $0.001\times 10^{-3}$).

We compare the $\epsilon_{\rm RS}^i$ values with the corresponding 
$N_{\rm RS}^i/N_{\rm RS}^{\rm tot}$ ratios.
The relative difference between the two is conservatively assigned as
the uncertainty on $\epsilon_{\rm RS}^i/\epsilon_{\rm WS}^i$,
although one expects at least part of the systematic uncertainty to
cancel in the efficiency ratio.
We reduce the six efficiency ratios simultaneously by
this uncertainty and repeat the $r_D$ calculation; we then increase the ratios by
this uncertainty, and again recalculate $r_D$.  
We quote the difference between the resulting $r_D$ values and the
default fit ($\pm 0.02
\times 10^{-3}$) as the systematic error due to an imperfect
description of the proper decay time distributions.

The systematic error due to the uncertainty in the associated signal fraction
is estimated by varying the fraction and repeating the fitting procedure.
Using the errors on the measured branching fractions \cite{PDG} of the
associated signal decay channels, we conservatively vary the amount of
associated signal 
 by $\pm 40\% $.
We recalculate $r_D$ and compare it to the default $r_D$ value; we
quote the difference, $0.002 \times  
10^{-3}$, as the systematic error from this source.

The quadratic sum of these individual contributions yields a
total systematic error of $\pm 0.14\times 10^{-3}$.

~\\
~\\
\centerline{\bf V. CONCLUSION}
~

In summary, we have searched for $D^0-\overline{D}{}^0$ mixing in
semileptonic $D^0$ decays. We independently measure the mixing
ratios $r_D^i$ in six $D^0$ proper decay time intervals.  By fitting a
constant to these six values we obtain the final result 
\begin{equation}
r_D=(0.20\pm 0.47 \pm 0.14)\times 10^{-3},
\label{eq_res2}
\end{equation}
where the first error is the statistical and the second the 
systematic uncertainty. As we do not observe a significant number of WS
$D^0$ decays, we obtain an upper limit for
the time-integrated mixing rate.  We use the Feldman-Cousins
approach \cite{FeldmanCousins} to calculate the upper limit in the
vicinity of the physics region boundary ($r_D\ge 0$). 
Using the result and the total error in Eq.~\ref{eq_res2}, this yields

\begin{equation}
r_D\le 1.0\times 10^{-3}~~{\rm at~90\%~C.L.}
\label{eq_ul2}
\end{equation}

\noindent This result is the most stringent experimental limit on the
time-integrated $D^0$ mixing rate obtained to date from 
semileptonic $D^0$ decays \cite{PDG}.

\begin{acknowledgments}
~\\
~\\

\centerline{\bf ACKNOWLEDGMENTS}

We thank the KEKB group for the excellent operation of the
accelerator, the KEK cryogenics group for the efficient
operation of the solenoid, and the KEK computer group and
the NII for valuable computing and Super-SINET network
support.  We acknowledge support from MEXT and JSPS (Japan);
ARC and DEST (Australia); NSFC (contract No.~10175071,
China); DST (India); the BK21 program of MOEHRD, and the
CHEP SRC and BR (grant No. R01-2005-000-10089-0) programs of
KOSEF (Korea); KBN (contract No.~2P03B 01324, Poland); MIST
(Russia); MHEST (Slovenia);  SNSF (Switzerland); NSC and MOE
(Taiwan); and DOE (USA).

\end{acknowledgments}


\begin{thebibliography}{99}
\bibitem{charge} Charge-conjugate modes are implied throughout this paper.
\end{thebibliography}
\end{document}